\documentclass[aip,
               amsmath,
               amssymb,
					author-year,
				   floatfix,
					reprint
					]{revtex4-1}
\usepackage{graphicx}

\DeclareMathOperator{\sign}{sign}
\DeclareMathOperator{\Ro}{Ro}

\begin{document}

\title{Dissipative inertial transport patterns near coherent
Lagrangian eddies in the ocean}

\author{Francisco J. Beron-Vera\footnote{Author to whom correspondence
should be addressed. Electronic mail: fberon@rsmas.miami.edu.}}
\affiliation{Department of Atmospheric Sciences, RSMAS, University
of Miami, Miami, Florida, USA}

\author{Mar\'{\i}a J. Olascoaga} \affiliation{Department of Ocean
Sciences, RSMAS, University of Miami, Miami, Florida, USA}

\author{George Haller} \affiliation{Institute for Mechanical Systems,
ETH Zurich, Zurich, Switzerland}

\author{Mohammad Farazmand} \affiliation{Department of
Mathematics, ETH Zurich, Zurich, Switzerland}

\author{Joaqu\'{\i}n Tri\~nanes\footnote{Also at Cooperative Institute
for Marine and Atmospheric Studies, RSMAS, University of Miami,
Miami, Florida and Instituto de Investigaci\'ones Tecnol\'oxicas,
Universidade de Santiago de Compostela, Santiago, Spain.}}
\affiliation{Physical Oceanography Division, AOML, NOAA, Miami,
Florida, USA}

\author{Yan Wang} \affiliation{Department of Ocean
Sciences, RSMAS, University of Miami, Miami, Florida, USA}

\date{Started: April 4, 2014. This version: \today.}

\begin{abstract}
  Recent developments in dynamical systems theory have revealed
  long-lived and coherent Lagrangian (i.e., material) eddies in
  incompressible, satellite-derived surface ocean velocity fields.
  Paradoxically, observed drifting buoys and floating matter tend
  to create dissipative-looking patterns near oceanic eddies, which
  appear to be inconsistent with the conservative fluid particle
  patterns created by coherent Lagrangian eddies.  Here we show
  that inclusion of inertial effects (i.e., those produced by the
  buoyancy and size finiteness of an object) in a rotating
  two-dimensional incompressible flow context resolves this paradox.
  Specifically, we obtain that anticyclonic coherent Lagrangian
  eddies attract (repel) negatively (positively) buoyant finite-size
  particles, while cyclonic coherent Lagrangian eddies attract
  (repel) positively (negatively) buoyant finite-size particles.
  We show how these results explain dissipative-looking satellite-tracked
  surface drifter and subsurface float trajectories, as well as
  satellite-derived \emph{Sargassum} distributions.
\end{abstract}



\maketitle

\begin{quotation}
  Satellite-tracked drifting buoy trajectories and satellite-derived
  algal distributions are commonly used in oceanography to infer
  Lagrangian aspects of the surface ocean circulation.  At the same
  time, dynamical systems techniques applied to surface ocean
  velocities inferred from satellite altimetry reveal persistent
  coherent Lagrangian eddies.  Paradoxically, buoys and algae display
  dissipative-looking patterns in contrast to the conservative-looking
  coherent Lagrangian eddies.  Here we show that the dissipative
  patterns are due to inertial effects superimposed on the conservative
  fluid patterns produced by coherent Lagrangian eddies.
\end{quotation}  

\section{Introduction}

The work reported in this paper provides an explanation for the
dissipative behavior of drifting buoys and floating matter on the
ocean surface near coherent Lagrangian (i.e., material) eddies.
Such eddies impose conservative behavior on nearby fluid particles
in incompressible two-dimensional flows, which seems at ods with
the observed dissipative patterns.

\begin{figure}[h]
  \centering%
  \includegraphics[width=.45\textwidth]{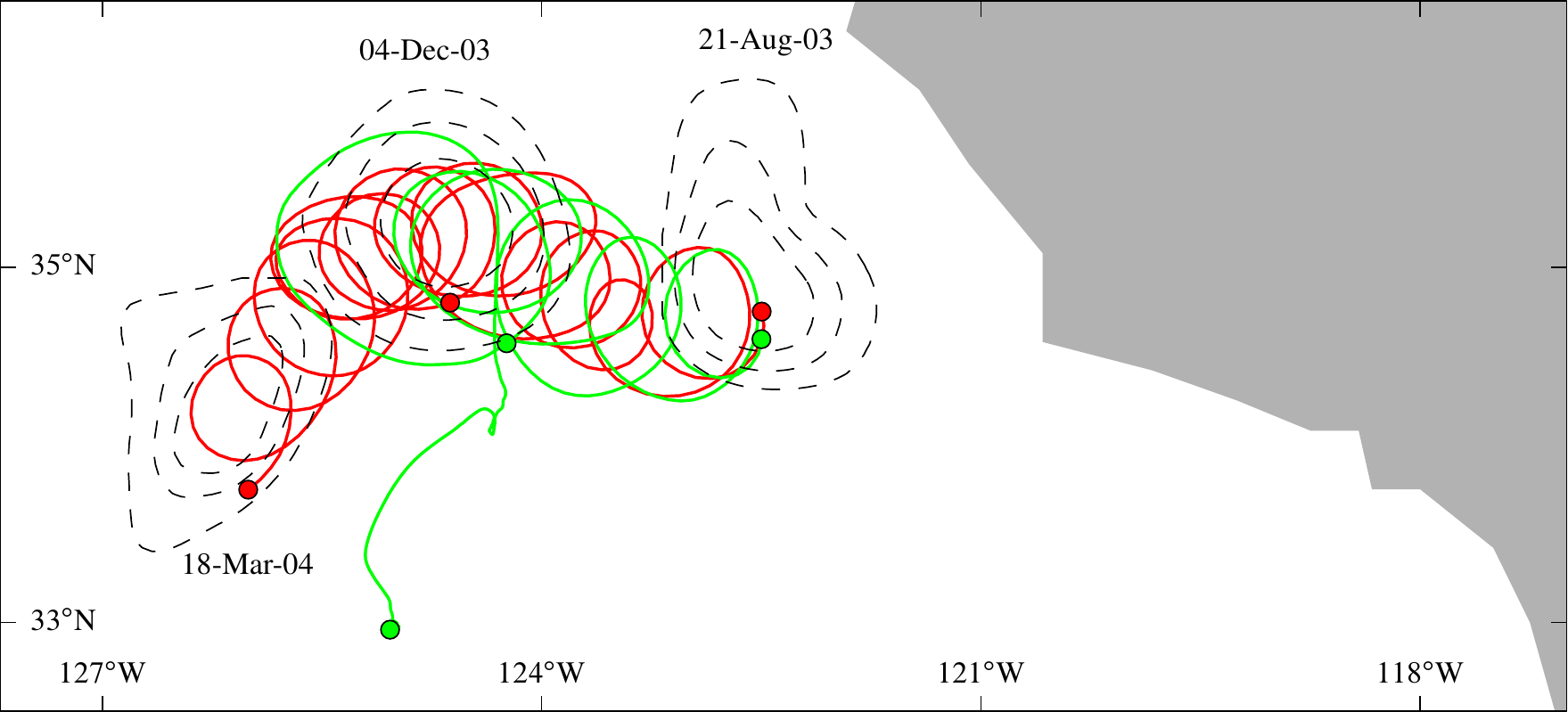}%
  \caption{Trajectories of two RAFOS floats (red and green curves)
  and selected snapshots of a westward propagating bulge of the
  satellite altimetric sea surface height (SSH) field (dashed curves
  indicate selected isolines) in the southeastern North Pacific.
  The dots indicate the positions of the floats on the dates that
  the SSH bulge is shown.}
  \label{fig:ssh}%
\end{figure}

A revealing example of observed dissipative behavior is that of two
RAFOS floats (acoustically tracked, subsurface, quasi-isobaric
drifting buoys) in the southeastern North Pacific (Fig.\ \ref{fig:ssh}).
Initially close together, the two floats (indicated in red and green
in Fig. 1) take significantly divergent trajectories on roughly the
same depth level (320 m) relative to the floats’ positional
uncertainty, which does not exceed 10 km \citep{Garfield-etal-99,
Collins-etal-13}.  This behavior at first sight might be attributed
to sensitive dependence of fluid particle trajectories on initial
particle positions in a turbulent ocean.  But analysis of satellite
altimetry measurements reveals that the floats on the date of closest
proximity fall within a region of roughly 100-km radius characterized
by a bulge of the sea surface height (SSH) field (selected isolines
are indicated by dashed curves in Fig.\ \ref{fig:ssh}).  This SSH
bulge propagates westward at a speed slower than the geostrophically
inferred clockwise tangential speed at its periphery, suggesting
the presence of a mesoscale anticyclonic eddy capable of holding
fluid \citep{Chelton-etal-11a}.  Indeed, this SSH eddy may be
identified with the surface manifestation of a California Undercurrent
eddy; such eddies, referred to as ``cuddies,'' have been argued to
be important transport agents \citep{Garfield-etal-99}.  However,
while one float is seen to loop anticyclonically accompanying the
eddy very closely, the other float anticyclonically spirals away
from the eddy rather quickly, representing a puzzle.

Even more puzzling is that the two floats actually initially lie,
as we show below, within the same \emph{coherent Lagrangian eddy}
\citep{Beron-etal-13, Haller-Beron-13, Haller-Beron-14}.  Representing
an elliptic Lagrangian coherent structure \citep[LCS; cf.][]{Haller-14},
the boundary of such an eddy defies the exponential stretching of
typical material loops in turbulence.  In effect, the eddy in
question exhibits minimal filamentation and deformation over several
months, and thus is expected to trap and carry within both floats.

The behavior of one of the floats supports this scenario and thus
the altimetry-derived (i.e., geostrophic) upper-ocean current
representation that sustains the coherent Lagrangian eddy.  Ageostrophic
processes of various types may be acting in the upper ocean, but
these cannot explain the substantively different behavior of the
other float.  Indeed, ageostrophic effects cannot be so different
on two initially nearby fluid particles in a region of mostly regular
flow.  Therefore, to resolve the puzzle, effects of a different
class must be accounted for.

Here we argue theoretically, and show both numerically and
observationally, that such effects can be produced by \emph{inertia},
i.e., buoyancy and size finiteness.  Inertial effects are commonly
considered in atmospheric transport studies.  These range from
studies aimed at explaining observed motion of meteorological
balloons \citep{Paparella-etal-97, Provenzale-99, Dvorkin-etal-01}
and spread of volcanic ash \citep{Haszpra-Tel-11}, to theoretical
and numerical studies of particulate matter dispersal
\citep{Haller-Sapsis-08, Sapsis-Haller-09, Tang-etal-09}. In
oceanography inertial effects have been also taken into account in
several problems including sedimentation \citep{Nielsen-94}, and
plankton sinking \citep{Stommel-49}, patchiness \citep{Reigada-etal-03}
and selfpropulsion \citep{Peng-Dabiri-09}.  However, they have been
rarely considered in the motion of drifting buoys, macroscopic
algae, or debris.  To the best of our knowledge, their potential
importance in influencing the motion of floats was only noted by
\citet{Tanga-Provenzale-94}.

Our theoretical results reveal that while the boundary of a coherent
Lagrangian eddy represents a transport barrier for fluid particles,
it does not do so for \emph{inertial particles}.  Instead, a coherent
Lagrangian eddy attracts or repels initially close inertial particles,
depending on the particles' density ratio with the ambient fluid
and on the polarity (rotation sense) of the eddy.

We first validate these numerically using altimetry-derived currents
in several regions of the ocean.  Next, we use our findings to
explain observed behavior in various ocean areas starting with the
aforementioned floats, then proceeding with satellite-tracked surface
drifting buoys (drifters), and finally macroscopic algae
(\emph{Sargassum}) distributions.

We emphasize that because our approach uses observationally-based
velocity, it enables feature matching and analysis of specific
measurements.  Furthermore, our approach is self-consistent within
the realm of incompressible two-dimensional flows.  This is in
marked contrast with a previous approach to surface ocean pattern
formation \citep{Zhong-etal-12}, which considered passive advection
by the surface velocity from a primitive-equation model (i.e., a
truncation of the three-dimensional velocity). This is destined to
create dissipative-looking patterns, but no actual passive tracer
follows such a virtual velocity field.

We also note that our results are not applicable to the problem of
accumulation of debris in subtropical gyres, which has been recently
investigated by \citet{Froyland-etal-14} using probabilistic methods.
The so-called great ocean garbage patches are produced by convergent
wind-induced Ekman transport \citep{Maximenko-etal-11}.  The Ekman
dynamics governing basin-scale motions are very different than the
quasigeostrophic dynamics governing mesoscale motions, our focus
here. The former can produce dissipative patterns on the surface
ocean by themselves, but the latter cannot unless inertial effects
are taken into account, as we noted above and demonstrate below.

The remainder of the paper is organized as follows.  Section 2.1
presents the mathematical setup required to formally introduce the
coherent Lagrangian eddy notion, which is briefly reviewed in Section
2.2.  The theoretical results relating to behavior of inertial
particles near coherent Lagrangian eddies are presented in Section
2.3.  In Section 2.4 further insight into inertial particle motion
is provided.  Numerical validation of the theoretical results is
presented in Section 3.  In Section 4 the theoretical results are
used to explain observed behavior in the ocean.  A summary and
discussion is offered in Section 5. Finally, Appendix \ref{app:rmr}
includes details of the asymptotic analysis leading to our theoretical
results, Appendix \ref{app:data} is reserved for the description
of the several datasets employed, and Appendix \ref{app:numerics}
gives some details of the various numerical computations performed.

\section{Theory}

\subsection{Mathematical setup}

We consider an incompressible two-dimensional velocity field,
$v(x,t)$, where position $x$ ranges on some open domain of
$\mathbb{R}^2$ and time $t$ is defined on a finite interval.
Specifically, we consider
\begin{equation}
  v = \frac{g}{f}\nabla^\perp\eta,
  \label{eq:v}
\end{equation}
where $\eta(x,t)$ is the SSH; the constants $f$ and $g$ stand for
Coriolis parameter (twice the local vertical component of the Earth's
angular velocity) and acceleration of gravity, respectively; and
$\perp$ represents a 90$^{\circ}$-anticlockwise rotation.  The
velocity field \eqref{eq:v} is representative of quasigeostrophic
motions in the upper ocean, i.e., characterized by a small Rossby
number, $\Ro := V/L|f|$, where $L$ and $V$ are typical length and
velocity scales, respectively. In particular, \eqref{eq:v} is
suitable to investigate transport near mesoscale eddies, our focus
here.  Fluid particles evolve according to
\begin{equation}
  \dot{x} = v.
  \label{eq:dxdt}
\end{equation}

Let $F_{t_0}^t(x_0) := x(t;x_0,t_0)$ be the flow map that takes
time $t_0$ positions to time $t$ positions of fluid particles obeying
\eqref{eq:dxdt}.  An objective (i.e., frame-invariant) measure of
material deformation in \eqref{eq:dxdt} is the right Cauchy--Green
strain tensor,
\begin{equation}
  C := (\mathrm{D}F)^\top \mathrm{D}F,
  \label{eq:C}
\end{equation}
where $\mathrm{D}$ stands for differentiation with respect to $x_0$.
For any smooth $v$, $F$ represents a diffeomorphism, which ensures
invertibility of $\mathrm{D}F$ and thus positive definiteness of
$C$. Furthermore, incompressibility of $v$ implies $\det C = 1$.
Consequently, eigenvalues $\{\lambda_i\}$ and normalized eigenvectors
$\{\xi_i\}$ of $C$ satisfy
\begin{equation}
  0 < \lambda_1 \leq \lambda_2 \equiv
  \frac{1}{\lambda_1},\quad 
  \xi_i \cdot \xi_j = \delta_{ij}\quad
  i,j = 1,2.
  \label{eq:eig}
\end{equation}

\subsection{Coherent Lagrangian eddies}

\begin{figure}[h]
  \centering%
  \includegraphics[width=.35\textwidth]{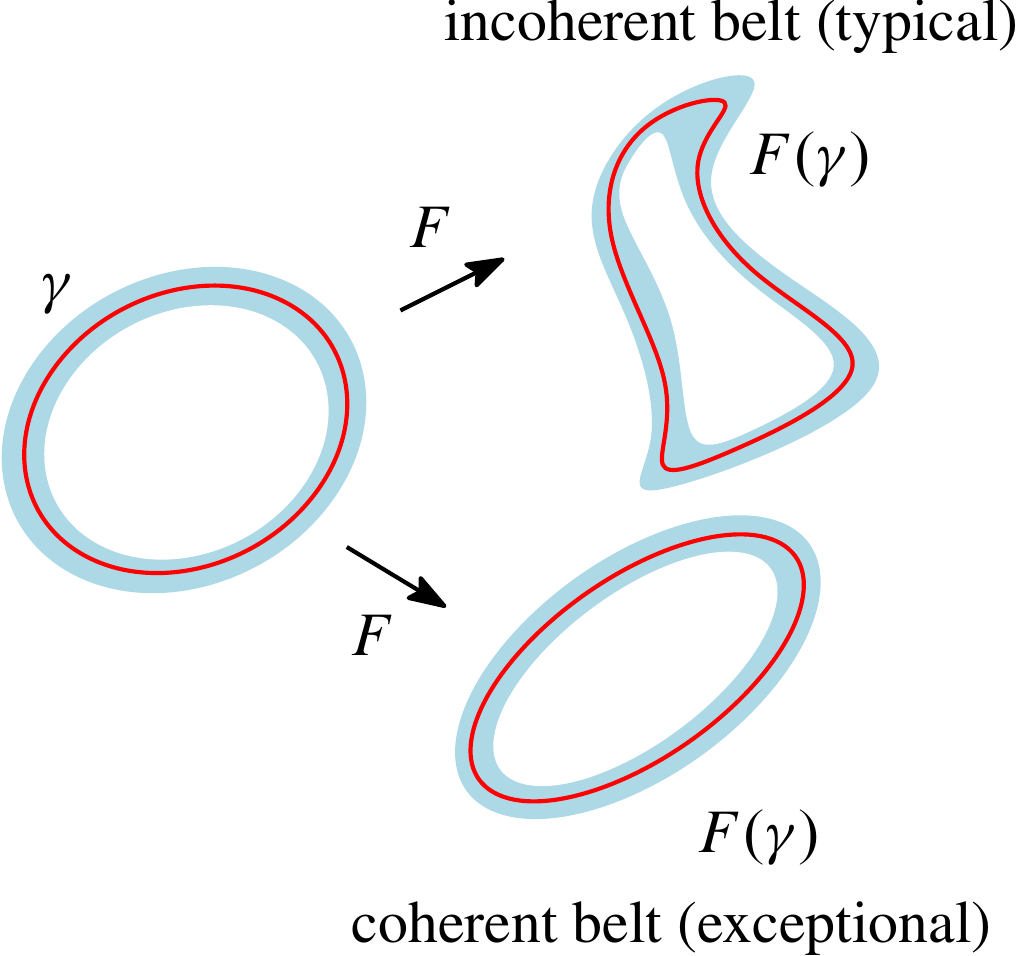}%
  \caption{A closed material curve $\gamma$ (red) at time $t_0$ is
  advected by the flow into $F(\gamma)$ at time $t$. The advected
  curve remains coherent if a thin material belt around it (light
  blue) shows no leading-order variation in averaged stretching after
  advection.}
  \label{fig:belts}%
\end{figure}

\citet{Haller-Beron-13} seek elliptic LCS as material loops with
small annular neighborhoods showing no leading-order variation in
averaged material stretching (Fig.\ \ref{fig:belts}).

Solving this variational problem reveals that elliptic LCS are
uniformly stretching: any of their subsets are stretched by the
same factor $\lambda$ under advection by the flow from time $t_0$
to time $t$.  The time $t_0$ positions of $\lambda$-stretching
elliptic LCS turn out to be limit cycles of one of the following
two objective ODE for parametric curves $s \mapsto r(s)$:
\begin{equation}
  r' = 
  \sqrt{
  \frac
  {\lambda_2 - \lambda^2}
  {\lambda_2 - \lambda_1}
  }
  \,\xi_1 
  \pm
  \sqrt{
  \frac
  {\lambda^2 - \lambda_1}
  {\lambda_2 - \lambda_1}
  }
  \,\xi_2,  
  \label{eq:eta}
\end{equation}
where the prime stands for $s$ differentiation.  More geometrically,
limit cycles of \eqref{eq:eta} are closed null geodesics of the
metric tensor $C - \lambda^2\mathrm{Id}$, which is Lorentzian in
the domain satisfying $\lambda_1 < \lambda^2 < \lambda_2$. This
provides a relativistic interpretation of coherent Lagrangian eddies
\citep[for details, cf.][]{Haller-Beron-13, Haller-Beron-14}.

The limit cycles of \eqref{eq:eta} will either grow or shrink under
changes in $\lambda$, forming smooth annular regions of nonintersecting
loops. The outermost member of such a band of coherent Lagrangian
loops will be observed physically as the boundary of the coherent
Lagrangian eddy.  We refer to these maximal elliptic LCS as
\emph{coherent Lagrangian eddy boundaries}.

Limit cycles of  \eqref{eq:eta} tend to exist only for $\lambda
\approx 1$.  Material loops characterized by $\lambda = 1$ resist
the universally observed material stretching in turbulence: they
reassume their initial arclength at time $t$.  This conservation
of arclength, along with the conservation of the enclosed area in
the incompressible case, creates extraordinary coherence for elliptic
LCS.

\subsection{Inertial effects near coherent Lagrangian eddies}

The Maxey--Riley equation \citep{Maxey-Riley-83} describes the
motion of inertial (i.e., buoyant, finite-size) particles, which
can deviate substantially from that of fluid (i.e., neutrally
buoyant, infinitesimally-small) particles
\citep[cf.][]{Cartwright-etal-10}.  Here we consider a simplified
version of the Maxey--Riley equation appropriate for inertial
particle motion in a quasigeostrophic flow. We further derive a
reduced form of this equation, which will allow us to assess behavior
near a coherent Lagrangian eddy.

Specifically, ignoring added mass effects, the Basset history term,
and so-called Faxen corrections, the Maxey--Riley equation for the
motion of a small spherical particle in the flow produced by
\eqref{eq:v} is given by
\begin{equation}
  \ddot{x} + f\dot{x}^\perp =
  \delta fv^\perp - \tau^{-1}\left(\dot{x} - v\right), 
  \label{eq:mr}
\end{equation}
where the constants
\begin{equation}
  \delta := \frac{\rho}{\rho^\mathrm{p}},\quad 
  \tau := \frac{2a^2}{9\nu\delta}. 
  \label{eq:dtatau}
\end{equation}
Here $\rho$ and $\nu$ are the fluid's density and viscosity,
respectively, and $\rho^\mathrm{p}$ and $a$ are the inertial
particle's density and radius, respectively.  The left-hand-side
of \eqref{eq:mr} is the inertial particle's absolute acceleration.
The first and second terms on the right-hand-side of \eqref{eq:mr}
are the flow force and Stokes drag, respectively.

The simplified form of the Maxey--Riley equation \eqref{eq:mr} was
priorly considered by \citet{Provenzale-99} with the following
differences.  First, the fluid relative acceleration, $\rho(\partial_t
v + v \cdot \nabla v)$, was included.  This term is one order of
magnitude smaller in Ro than $fv^\perp$ and thus is conveniently
neglected here.  Second, a centrifugal force term was included too,
but this is actually balanced by the gravitational force on the
horizontal plane.  Third, a vertical buoyancy force term was
considered, but this in the end played no role as the focus was on
motion on a horizontal plane, as here.

We introduce the small nondimensional parameter:
\begin{equation}
  \varepsilon := \tau\frac{V}{L} =
  \frac{2}{9\delta}\left(\frac{a}{L}\right)^2\mathrm{Re} =
  \frac{\mathrm{St}}{\delta} \ll 1,
  \label{eq:vep0}
\end{equation}
where $\mathrm{Re}$ and $\mathrm{St}$ are Reynolds and Stokes
numbers, respectively.  Consistent with the quasigeostrophic scaling
assumptions leading to the fluid velocity field \eqref{eq:v}, we
can set
\begin{equation}
  \varepsilon = O(\Ro).
  \label{eq:vep}
\end{equation}
In Appendix \ref{app:rmr} we show that inertial particle motion
characterized by \eqref{eq:vep}, e.g., the motion of particles much
smaller than the typical lengthscale of the flow, is controlled
at leading order by
\begin{equation}
  \dot{x} = v^\mathrm{p} =  v + \tau \left(\delta - 1\right)
  fv^\perp,
  \label{eq:rmr}
\end{equation}
which is the reduced form of the Maxey--Riley equation we shall
use.  This reduced equation is valid up to an $O(\varepsilon^2$)
error, after particles reach the vicinity of an attracting slow
manifold exponentially fast.

\begin{figure}[h]
  \centering%
  \includegraphics[width=.45\textwidth]{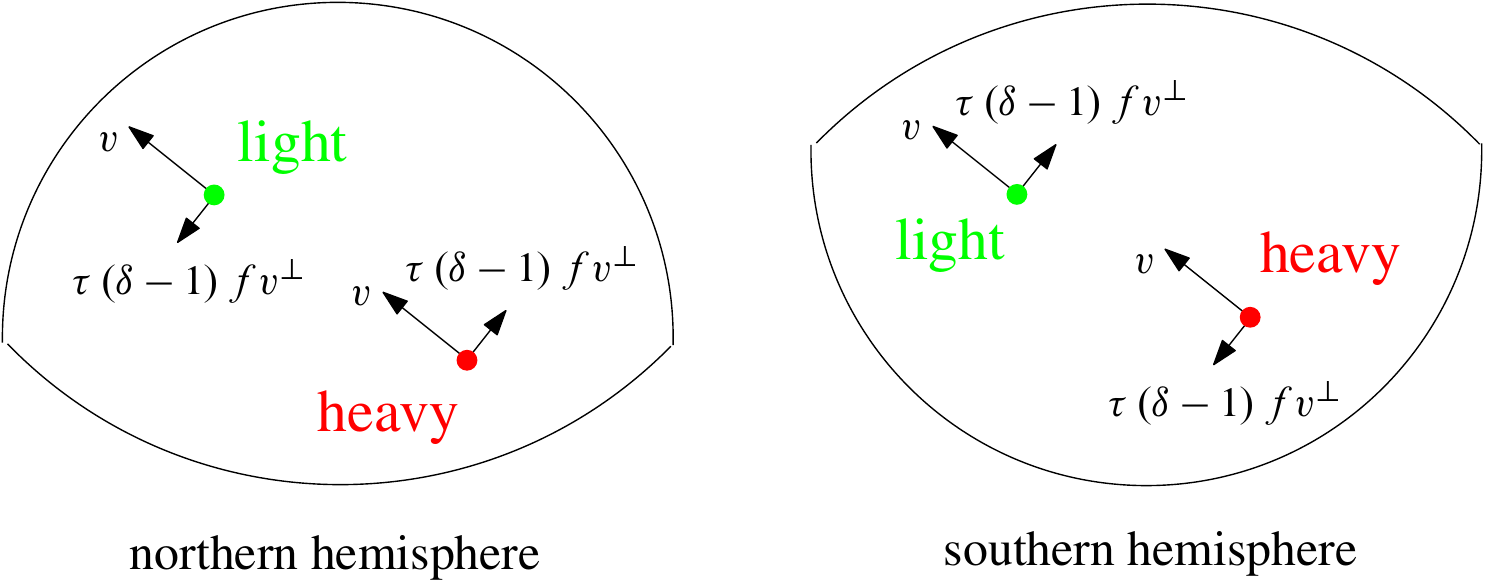}%
  \caption{Velocity contributions to inertial particle's
  velocity: light (heavy) particle motion deflects to the right (left)
  of fluid particle motion in the northern hemisphere and vice
  versa in the southern hemisphere.}
  \label{fig:vel}%
\end{figure}

Comparison of \eqref{eq:dxdt} and \eqref{eq:rmr} reveals that the
fluid velocity, $v$, and inertial particle velocity, $v^\mathrm{p}$,
differ by a dissipative $O(\varepsilon)$ term.  In the northern
hemisphere ($f > 0$) this term acts to deflect the motion of
positively buoyant ($\delta > 1$) finite-size or \emph{light}
particles to the left of the motion of fluid particles, while it
acts to deflect the motion of negatively buoyant ($\delta < 1$)
finite-size or \emph{heavy} particles to the right; in the southern
hemisphere ($f < 0$) it acts the opposite way (Fig.\ \ref{fig:vel}).

Inertial effects, therefore, should promote divergence away from,
or convergence into, coherent Lagrangian eddies when otherwise fluid
particles circulate around them.  

Specifically, let $\gamma$ be the boundary of a coherent Lagrangian
eddy at time $t$ and $U_\gamma$ the region $\gamma$ encloses. Up
to an $O(\varepsilon^2)$ error, the flux across $\gamma$ is given
by
\begin{eqnarray}
  \Phi_{\gamma} &=& \oint_{\gamma}\,(v - v^\mathrm{p}) \cdot
  \mathrm{d}x^\perp\nonumber\\ &=& \int_{U_\gamma} \nabla \cdot
  (v^\mathrm{p} - v)\,\mathrm{d}^2x\nonumber\\ &=& \tau(1 -
  \delta)f\int_{U_\gamma}\omega\,\mathrm{d}^2x,
  \label{eq:Phi}
\end{eqnarray}
where the loop integral is taken anticlockwise and $\omega := -\nabla
\cdot v^\perp (= gf^{-1}\nabla^2\eta)$ is the fluid's vorticity.
Inspection of expression \eqref{eq:Phi} leads to the following
conclusions:
\begin{enumerate}
 \item cyclonic ($f\omega > 0$) coherent Lagrangian eddies
	 attract ($\Phi_\gamma < 0$) light ($\delta
	 > 1$) particles and repel ($\Phi_\gamma
	 > 0$) heavy ($\delta < 1$) particles; while
 \item anticyclonic ($f\omega < 0$) coherent Lagrangian eddies
	 attract ($\Phi_\gamma < 0$) heavy ($\delta
	 < 1$) particles and repel ($\Phi_\gamma
	 > 0$) light ($\delta > 1$) particles.
\end{enumerate}

Our results concerning heavy particles confirm the numerical
observations of \citet{Provenzale-99} and extend them to the behavior
of light particles.

Our computations below are based on the reduced Maxey--Riley equation
\eqref{eq:rmr}, which we refer to as the \emph{inertial equation}.
This follows the terminology of \citet{Haller-Sapsis-08}, who
obtained the reduced form of a system similar to \eqref{eq:mr} in
a nonrotating frame.  Considering \eqref{eq:rmr} is advantageous
computationally and, as we will show, sufficiently accurate for the
verification of our theoretical results.

\subsection{Inertial Lagrangian Coherent Structures}

While motion of inertial particles is not constrained by LCS, it
is tied to analogous exceptional invariant curves referred to as
\emph{inertial LCS} \citep[or \emph{iLCS}; cf.][]{Haller-Sapsis-08}.

Of particular interest for our purposes here are hyperbolic iLCS
of attracting type.  These can be obtained by applying recent LCS
theory results \citep{Haller-Beron-12, Farazmand-Haller-13,
Farazmand-etal-14, Haller-14} on system \eqref{eq:rmr}.  Specifically,
iLCS at time $t_0$ which attract nearby inertial particle trajectories
over $[t_0,t]$ are invariant curves $s \mapsto r(s)$ that
satisfy 
\begin{equation}
  r' = \xi^\mathrm{p}_1
  \quad\text{or}\quad
  r' = \xi^\mathrm{p}_2
  \label{eq:xi}
\end{equation}
and
\begin{equation}
  \sqrt{\smash[b]{\lambda^\mathrm{p}_2}} > 1
  \quad\text{or}\quad
  \sqrt{\smash[b]{\lambda^\mathrm{p}_1}} < 1 
  \label{eq:rho}
\end{equation}
for $t < t_0$ or $t > t_0$, respectively. Here $\{\lambda_i^\mathrm{p}\}$
and $\{\xi_i^\mathrm{p}\}$ are eigenvalues and eigenvectors,
respectively, of the Cauchy--Green tensor, $C^\mathrm{p}$, derived
from system \eqref{eq:rmr}, which is an objective measure of
deformation in that system.  In forward time, segments of these
invariant lines squeeze and stretch, respectively.  As a result,
they can be referred to as \emph{inertial squeezelines} and
\emph{inertial stretchlines}, respectively.  In a similar manner
as the $\lambda$-lines discussed above, these invariant lines admit
a null geodesic interpretation.  In this case, the relevant Lorentzian
metric tensor is given by $C^\mathrm{p}\Omega - \Omega C^\mathrm{p}$,
where $\Omega$ is a 90$^{\circ}$-anticlockwise rotation matrix
\citep{Farazmand-etal-14}.

\section{Simulations}

Here we present numerical results that confirm our theoretical
predictions for the motion of inertial particles near coherent
Lagrangian eddies in the ocean.  

In each of our numerical tests a coherent Lagrangian eddy was
detected assuming that fluid trajectories are governed by \eqref{eq:dxdt}
with the velocity field given in \eqref{eq:v}; the SSH field is
constructed using satellite altimetry measurements \citep{Fu-etal-10}.
All eddies were detected from 90-day forward integration and found
to have $\lambda = 1$.  Successive positions of the boundaries of
the eddies past the detection time were obtained from advection.
Inertial particles were assumed to have $a = 0.25$ m, which is a
realistic radius value for commonly employed spherical drifting
buoys.  Both light and heavy particles were considered, with $\delta
= 1.1$ and $0.9$, respectively.  These density ratio values are
generally representative of surface floating and slowly or sinking
buoys, respectively.  For typical oceanic mesoscale eddies, with
diameter $L \sim 150$ km and tangential velocity at the boundary
$V \sim 0.1$ m s$^{-1}$, these inertial particle parameter choices
give $\varepsilon \sim 0.01$.  This $\varepsilon$ value turned out
to be small enough for particle motion obeying the Maxey--Riley
equation \eqref{eq:mr} to exhibit behavior qualitatively similar
to that satisfying the inertial equation \eqref{eq:mr} employed in
our simulations.

\begin{figure}[h]
  \centering%
  \includegraphics[width=.45\textwidth]{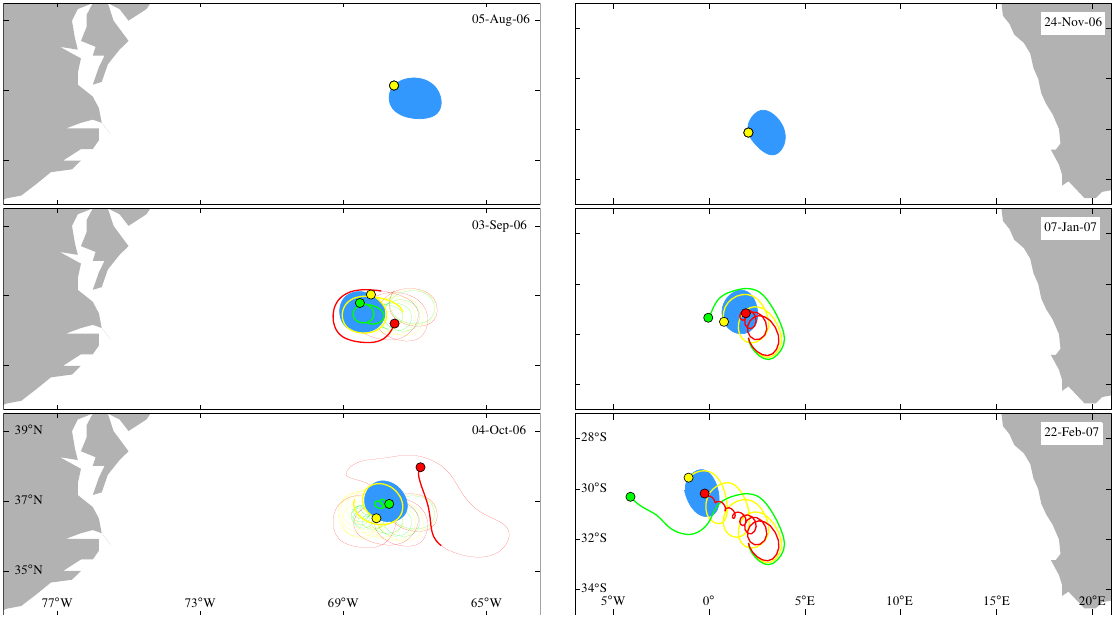}%
  \caption{Simulated trajectories of light (green), heavy (red),
  and fluid (yellow) particles initially on the boundaries of two
  mesoscale coherent Lagrangian eddies (light blue) extracted from
  altimetry-derived velocity.  Advection for the fluid particles
  is supplied by the altimetry-derived velocity, and heavy and light
  particle motion is controlled by the inertial equation \eqref{eq:rmr}.
  The eddies are identifiable with a cyclonic cold-core Gulf Stream
  ring (left panel) and an anticyclonic Agulhas ring (right panel).}
  \label{fig:sim}%
\end{figure}

We begin by discussing the results of tests involving light and
heavy particles initially located on the same position on the
boundary of a coherent Lagrangian eddy. The results are summarized
in Fig.\ \ref{fig:sim}, which consider a cyclonic (left panel) and
an anticyclonic (right panel) eddy, both indicated in light blue.
The eddy in the left panel is identifiable with an Agulhas ring,
while that in the right panel with a cold-core Gulf Stream ring.
The arclength of the boundary of each eddy on the detection date
is reassumed 90 days after (recall that the eddies have $\lambda =
1$).  Coherence is nevertheless observed well beyond 90 days
consistent with previous analyses of the satellite altimetry dataset
\citep{Beron-etal-13, Haller-Beron-13}.  This is evident from the
complete absence of filamentation. The light (green) and heavy (red)
particles behave quite differently than the fluid particle (yellow)
initially lying on the same position as the inertial particles on
the boundary of each eddy. The fluid particle remains on all dates
shown on the boundary of the Lagrangian eddy carrying the particle.
Consistent with our predictions, the light (heavy) particle spirals
out of (into) the Agulhas ring, while it spirals into (out of) the
cold-core Gulf Stream ring.

\begin{figure}[h]
  \centering%
  \includegraphics[width=.45\textwidth]{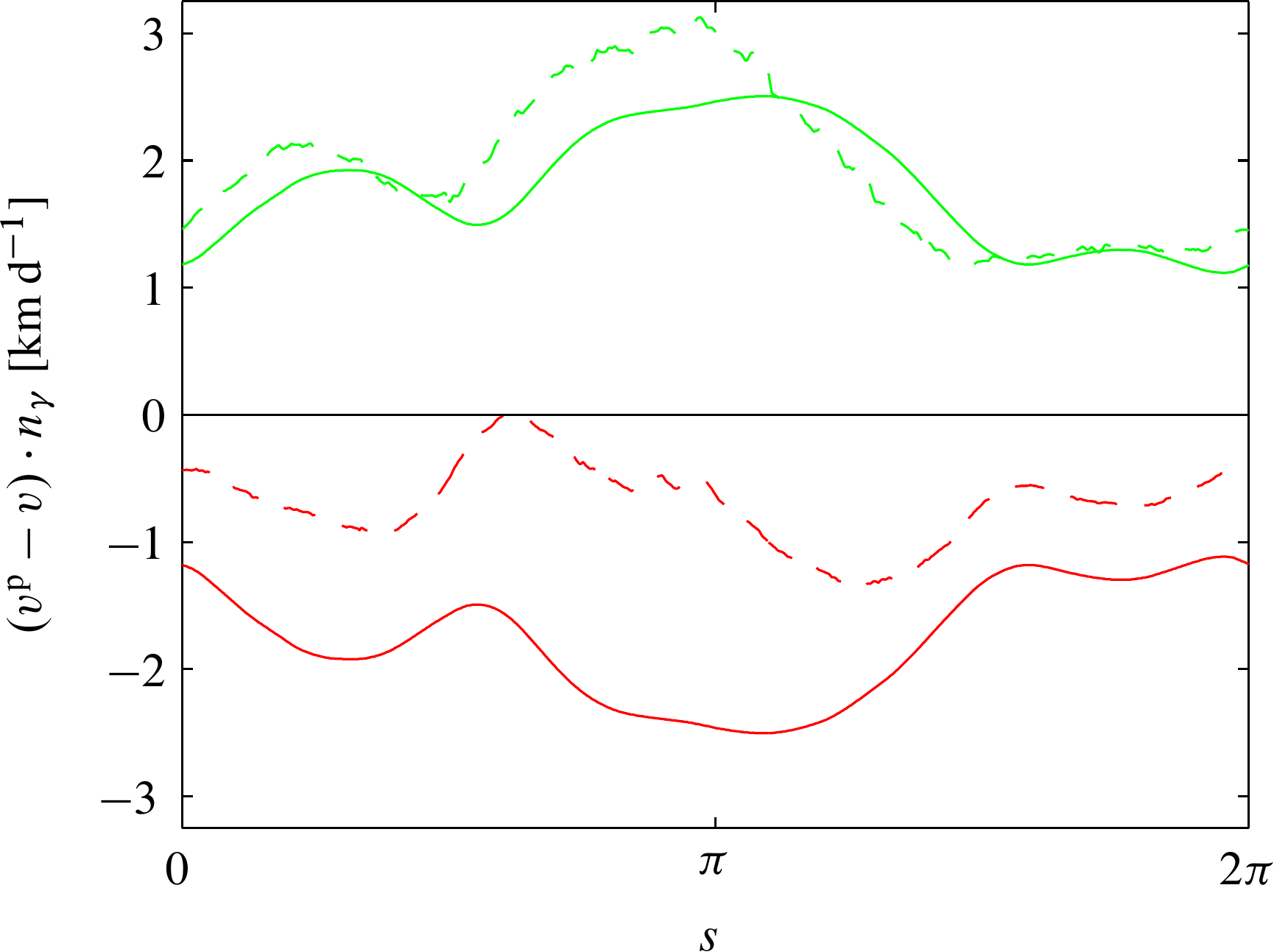}%
  \caption{Pointwise flux on 24 November 2006 of simulated light
  (green) and heavy (red) particles across the boundary of the
  anticyclonic coherent Lagrangian eddy identified in the previous
  figure as an Agulhas ring. Solid and dashed curves correspond to
  simulations based on the inertial \eqref{eq:rmr} and Maxey--Riley
  \eqref{eq:mr} equations, respectively.}
  \label{fig:flx}%
\end{figure}

We now provide more explicit support to our predictions by presenting
the results from the computation of the pointwise flux of inertial
particles across the boundary of a coherent Lagrangian eddy.  Across
a material loop $\gamma$, the pointwise flux of inertial particles
is given by $(v^\mathrm{p} - v) \cdot n_\gamma$, where $n_\gamma$
is the outer unit normal to $\gamma$.  Taking $\gamma$ as the
boundary of the eddy identified above as an Agulhas ring, the latter
is plotted in Fig.\ \ref{fig:flx} on 24 November 2006 as a function
of the boundary parameter $s$, chosen to be an azimuthal angle.
The pointwise fluxes of light (solid green) and heavy (solid red)
particles are everywhere inward and outward, respectively, along
the boundary of the anticyclonic eddy in question.  Thus our sign
predictions for the total flux extend to the pointwise flux in this
example.

\begin{figure}[h]
  \centering%
  \includegraphics[width=.451\textwidth]{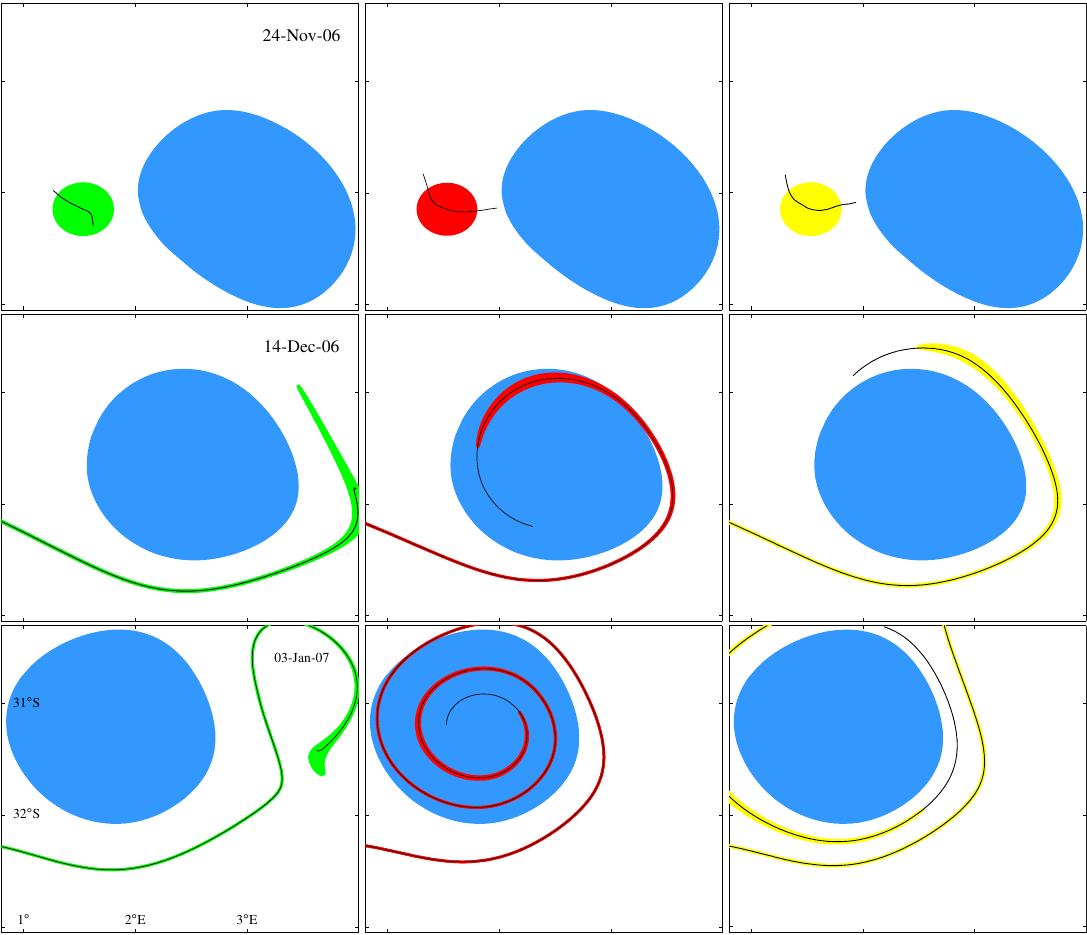}%
  \caption{Simulated evolution of patches of light (green), heavy
  (red), and fluid (yellow) particles initially outside of the
  coherent Lagrangian Agulhas ring in the previous figures (light
  blue).  Centerpiece attracting iLCS and LCS for the inertial and
  fluid particle patches, respectively, are indicated in black.
  Advection for fluid the particles is supplied by altimetry-derived
  velocities, and inertial particle motion is controlled by
  \eqref{eq:rmr}.}
  \label{fig:sim-ilcs}%
\end{figure}

We now turn to illustrate in Fig.\ \ref{fig:sim-ilcs} that the
evolution of inertial particles is tied to attracting iLCS. This
is done for patches of light (green) and heavy (red) particles lying
initially outside the coherent Lagrangian Agulhas ring discussed
above (light blue).  Shown attracting iLCS (black) were computed
as most stretching inertial stretchlines through each patch.  This
was done on the eddy detection time from a 90-day-forward integration.
The evolution of each inertial stretchline was determined by
advection.  After experiencing substantial stretching the heavy
particle patch is repelled away from the eddy.  By contrast, the
light particle patch spirals into the eddy.  As expected, the
attracting iLCS forms the centerpiece of the patch in each case.
For completeness, the evolution of a fluid patch (yellow) is also
shown.  In this case, too, the patch evolution is tied to its
centerpiece attracting LCS, computed also as the most stretching
stretchline through patch.  Consistent with the material nature of
the boundary of the coherent eddy, the fluid patch spirals around
the eddy without penetrating it.

\begin{figure}[h]
  \centering%
  \includegraphics[width=.45\textwidth]{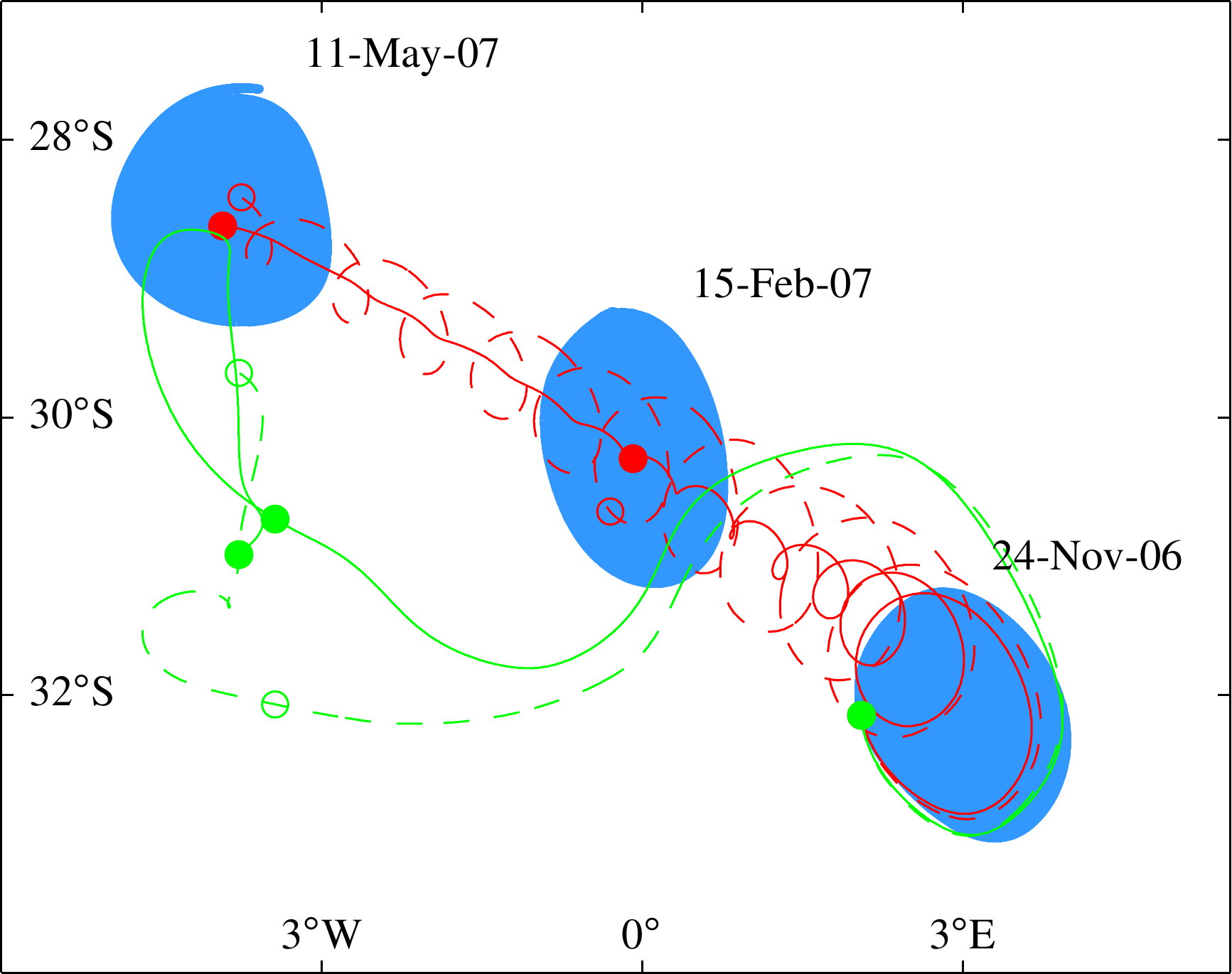}%
  \caption{Simulated trajectories of light (green) and heavy (red)
  particles initially lying on the same location on the boundary of
  the coherent Lagrangian Agulhas ring of the previous figures (light
  blue) based on inertial (solid) and Maxey--Riley (dashed) equations.}
  \label{fig:trj}%
\end{figure}

Finally, we show that for the parameters chosen, the qualitative
behavior near a coherent Lagrangian eddy described by the Maxey--Riley
\eqref{eq:mr} equation is captured by the inertial equation
\eqref{eq:rmr}.  This is illustrated in Fig.\ \ref{fig:flx}, where
the dashed lines correspond to pointwise flux calculations based
on the Maxey--Riley equation.  This calculation involves
trajectories started earlier, with a small (10\% of the fluid
velocity) perturbation to the velocity given in \eqref{eq:rmr}.
While strict convergence of inertial- and Maxey--Riley-equation-based
flux calculations has not been attained after 30 days of integration,
both flux calculations agree in sign and share a similar structure.
The slow convergence to the inertial manifolds arises from the
highly unsteady nature of the altimetry-derived flow.  Under such
conditions, pronounced convergence is only observable near sufficiently
persistent attracting sets.  This is illustrated in Fig.\ \ref{fig:trj},
which shows trajectories of light (green) and heavy (red) particles
lying on 24 November 2006 at the same position on the boundary of
the coherent Lagrangian Agulhas ring considered in the pointwise
flux calculations.  As in Fig.\ \ref{fig:flx}, solid and dashed
curves correspond to calculations based on inertial and Maxey--Riley
equations, respectively. Once again, while details of inertial- and
Maxey--Riley-equation-based trajectories are different and convergence
can only be expected when particles are heavy, our predictions are
seen to hold well: the heavy particle is attracted by the anticyclonic
coherent Lagrangian eddy in question, whereas the light particle
is repelled away from it.

\section{Observations}

In this section we discuss four sets of ocean observations that can
be explained using our predictions for the motion of inertial
particles near mesoscale coherent Lagrangian eddies.

\begin{figure}[h]
  \centering%
  \includegraphics[width=.45\textwidth]{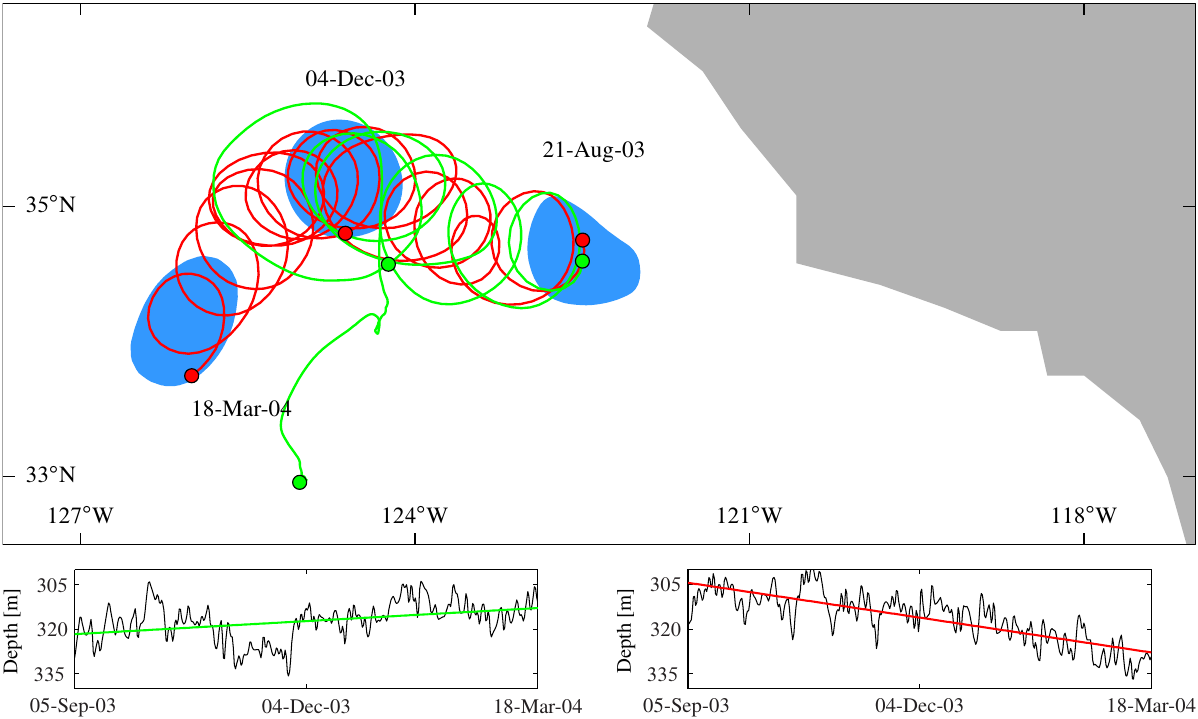}%
  \caption{(top panel) Trajectories of the two RAFOS floats in Fig.\
  1 (green and red curves) and snapshots of an anticyclonic coherent
  material eddy detected from altimetry (light blue). The dots
  indicate the positions of the floats on the dates that the eddy
  is shown.  (bottom-left panel) Depth of the green float as a
  function of time (black) with linear trend indicated (green).
  (bottom-right panel) As in the bottom-left panel, but for the red
  float.}
  \label{fig:rafos2}%
\end{figure}

The first set of observations, discussed in the Introduction, concern
two RAFOS floats in the southeastern North Pacific. The floats took
divergent trajectories despite their initial proximity within an
anticyclonic mesoscale eddy. This eddy was revealed from the Eulerian
footprints in the altimetric SSH field of a California Undercurrent
eddy or ``cuddy.''  As we anticipated in the Introduction and now
explicitly show in the top panel of Fig.\ \ref{fig:rafos2}, the
altimetry-derived velocity field sustains a coherent Lagrangian
eddy in nearly the same position as the SSH eddy.  The eddy, obtained
from a 90-day-forward integration with $\lambda = 1$, is depicted
(in light blue) on the detection date and two subsequent dates.
The trajectory of each of the two floats is indicated by a curve,
with their initial position highlighted by a dot. The divergent
behavior of the float trajectories can be explained by inertial
effects as follows.  Note that the float indicated in green takes
a slightly ascending trajectory (Fig.\ \ref{fig:rafos2}, bottom-left
panel), whereas the float indicated in red takes a slightly descending
trajectory (Fig.\ \ref{fig:rafos2}, bottom-right panel).  Thus the
ascending float represents a positively buoyant (i.e., light) object,
whereas the descending float represents a negatively buoyant (i.e.,
heavy) object.  The radius of the looping trajectory taken by the
ascending float is seen to increase as the float drifts westward
accompanying the anticyclonic coherent Lagrangian eddy revealed
from altimetry, which is eventually abandoned by the float.  This
behavior adheres closely to what we have predicted for light
particles.  The radius of the looping trajectory taken by the
descending float is not seen to decrease as the float drifts westward
accompanying the eddy in question.  However, in marked contrast
with the ascending float, the descending float remains within this
eddy.  This behavior adheres to what we have predicted for heavy
particles near anticyclonic coherent Lagrangian eddies.

\begin{figure}[h]
  \centering%
  \includegraphics[width=.45\textwidth]{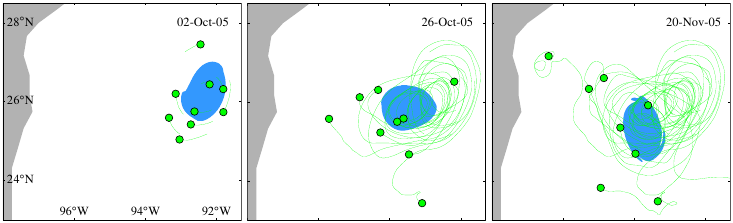}%
  \caption{Trajectories of satellite-tracked surface drifters (green
  curves) and snapshots of a coherent Lagrangian Loop Current ring
  detected from altimetry (light blue).  The dots indicate the
  positions of the drifters on the corresponding date.}
  \label{fig:rita}%
\end{figure}

The second set of observations concerns trajectories of satellite-tracked
surface drifters deployed in the Gulf of Mexico ahead of hurricane
Rita in September 2005.  The drifters were deployed inside a Loop
Current ring detected from its Eulerian footprints in the altimetric
SSH field.  A 30-day-forward integration of the altimetry-derived
velocity field reveals that the anticyclonic SSH eddy contains a
$\lambda = 1$ coherent Lagrangian eddy with a radius of roughly 100
km, about 25-km smaller than that of the approximately circular
area occupied by the SSH bulge.  Figure \ref{fig:rita} shows the
coherent Lagrangian eddy (light blue) on the detection date and two
subsequent positions obtained from advection.  The trajectories of
the drifters (a total of nine) are depicted in green (positions on
the date shown are indicated by dots).  Hurricane Rita made landfall
about one week prior to the detection date, so neither the altimetry
signal nor the motion of the drifters are affected by the high winds
associated with this tropical cyclone system.  Three drifters lie
inside the eddy on the detection date, while the remaining six are
located outside of the eddy, but close by its boundary.  Overall,
the drifters undergo growing looping trajectories.  More than one
month after the detection date, all nine drifters are found away
from the center of the eddy, with three lying on its border and six
lying well away from it. Noting that the drifters maintain afloat
on the ocean surface, this behavior can be expected, given that
anticyclonic coherent Lagrangian eddies repel away light particles
according to our results.

\begin{figure}[h]
  \centering%
  \includegraphics[width=.45\textwidth]{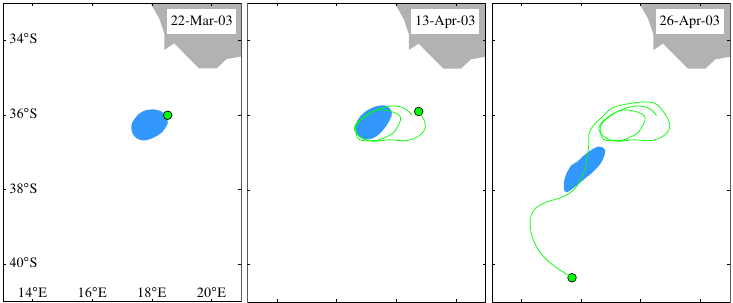}%
  \caption{Trajectory of an Argos-tracked surface drifter (green
  curve) and snapshots of a short-lived coherent Lagrangian Agulhas
  ring detected from altimetry (light blue).  A dot indicates
  the position of the drifter on the corresponding date.}
  \label{fig:astrid}%
\end{figure}

The third set of observations involves the trajectory of a surface
drifters tracked by the Argos satellite system, which was deployed
inside an Agulhas ring, named Astrid, as part of the Mixing of
Agulhas Rings Experiment (MARE) \citep{vanAken-etal-03}.  (Two
additional drifters were deployed during MARE whose trajectories
have not been possible to access.  However, all three drifters
behave similarly as it can be seen in Fig.\ 6 of \citet{vanAken-etal-03}.)
Detected from its Eulerian footprints in the altimetric SSH field,
ring Astrid was subjected to a detailed survey.  Hydrographic casts
across ring Astrid indicated the presence of a warm and saline core.
Acoustic Doppler current profiling revealed that Astrid had, in
addition to the baroclinic flow around its core, a significant
barotropic component.  A 30-day forward integration of the
altimetry-derived velocity field reveals a coherent Lagrangian eddy
with $\lambda = 1$.  This eddy has a mean radius of roughly 100 km,
about half that of the approximately circular region spanned by the
SSH bulge.  Figure \ref{fig:astrid} shows selected snapshots of the
coherent Lagrangian eddy (light blue) on three dates starting from
the detection date.  The trajectory of the Argos-tracked surface
drifter is indicated in green (dots indicate the positions of the
drifter on the corresponding dates).  The drifter is seen to develop
counterclockwise looping trajectories.  This grows in radius and
quite quickly the drifters abandon the vicinity of the eddy.  The
coherent Lagrangian eddy is rather short lived, thereby not revealing
the presence of a well-developed Agulhas ring (possibly consistent
with the lack of a well-defined core in the in-situ velocity
measurements).  However, over the lifespan of the eddy, the drifters'
behavior is consistent with our predictions for a light particle.
Therefore, our results offer an explanation for its motion.

\begin{figure}[h]
  \centering%
  \includegraphics[width=.45\textwidth]{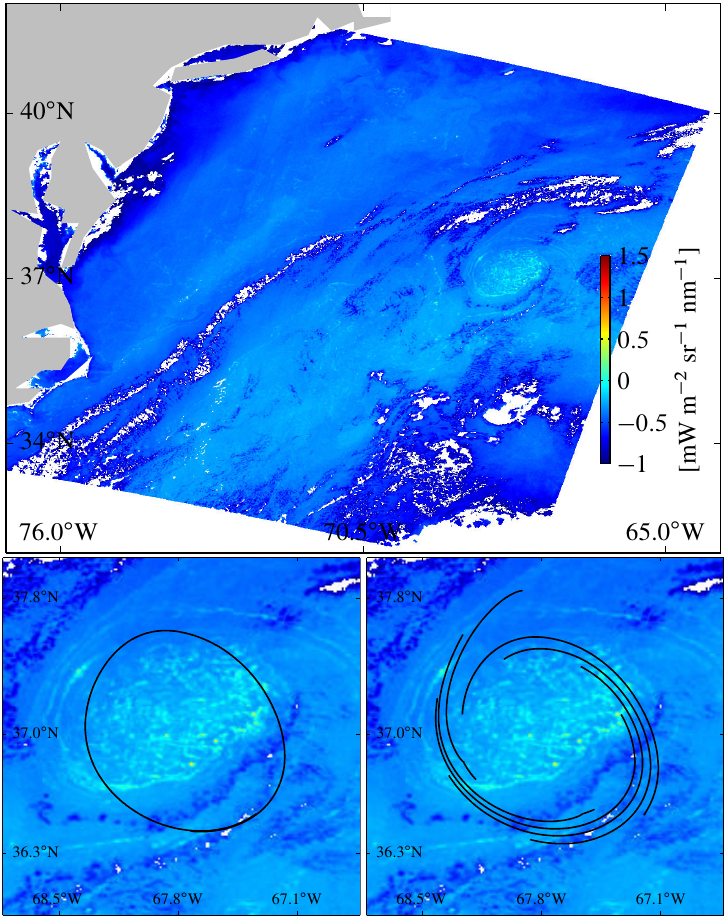}%
  \caption{(top panel)  Maximum Chlorophyll Index (MCI)  in the
  western North Atlantic inferred from the Medium Resolution Imaging
  Spectrometer (MERIS) aboard \emph{Envisat} on 4 October 2006.
  Floating \emph{Sargassum} corresponds to MCI values in excess of
  $-0.25$ mW m$^{-2}$ sr$^{-1}$ nm$^{-1}$.  (bottom-left panel)
  Boundary of a coherent Lagrangian cold-core Gulf Stream ring
  detected from altimetry (black) overlaid on the \emph{Sargassum}
  distribution.  (bottom-right panel) Altimetry-based attracting
  light iLCS (black) overlaid on the \emph{Sargassum} distribution.}
  \label{fig:sarga}%
\end{figure}

Finally, the fourth set of observations involves distribution of
floating \emph{Sargassum} on the sea surface in the western North
Atlantic inferred from the Medium Resolution Imaging Spectrometer
(MERIS) aboard \emph{Envisat} (Fig.\ \ref{fig:sarga}, top panel).
\emph{Sargassum} corresponds to Maximum Chlorophyll Index (MCI)
values exceeding $-0.25$ mW m$^{-2}$ sr$^{-1}$ nm$^{-1}$.  Detected
on 4 October 2006, the feature of interest takes a spiraled shape
and lies inside a coherent Lagrangian cold-core Gulf Stream ring
as revealed from altimetry.  In the bottom-left panel of Fig.\
\ref{fig:sarga} the material boundary of this cyclonic coherent
ring is shown overlaid on the \emph{Sargassum} feature in question.
This was obtained in Section 3 from advection of a coherent Lagrangian
eddy boundary computed on 5 August 2006 from a 90-day-forward
integration of the altimetry-derived velocity field (cf.\ Fig.\
\ref{fig:sim}, top-right panels).  The accumulation of the floating
\emph{Sargassum} inside the Gulf Stream ring is consistent with the
behavior of inertial particles near cyclonic coherent Lagrangian
eddies, which attract light particles according to our results.
The spiraled shape of the \emph{Sargassum} distribution inside the
ring is consistent too with the spiraled shape acquired by
altimetry-based attracting light iLCS (parameters are as in Section
3).  Selected iLCS are shown overlaid on the \emph{Sargassum}
distribution in the bottom-right panel of Fig.\ \ref{fig:sarga}.
These were obtained as backward-time light inertial squeezelines
initialized along the boundary of the Gulf Stream ring on the date
shown.  The direction of the spiraling inertial particle motion
along these iLCS is inward, as direct integration of the inertial
particle equation reveals.

\section{Summary and discussion}

In this paper we have provided an explanation for the observed
tendency of drifting buoys and floating matter on the surface of
the ocean to produce dissipative-looking patterns.  This resolves
an apparent paradox with the conservative-looking distributions
that tracers passively advected by a rotating two-dimensional
incompressible flow display. Our explanation takes into account
inertial effects, i.e., those produced by the buoyancy and size
finiteness of an object immersed in such a flow.  These are described
by a simplified Maxey--Riley equation consistent with a flow produced
by a quasigeostrophic velocity where the pressure field is entirely
due to differences in sea surface height.  Because the latter are
readily available from satellite altimetry measurements, our approach
enables feature matching and analysis of specific observations.
Furthermore, our approach is self-consistent within the realm of
two-dimensional incompressible flows.

We have found that anticyclonic coherent Lagrangian eddies attract
(repel) heavy (light) particles, while cyclonic ones behave the
opposite way.  We verified these results numerically using mesoscale
SSH fields constructed from satellite altimetry measurements in
various places of the ocean.  Our findings also explained
dissipative-type behavior shown by four sets of observations:
divergent motion of subsurface floats initially inside a California
Undercurrent eddy or ``cuddy;'' dispersion of surface drifters away
from a Loop Current ring; ejection of surface drifters out of a
well-studied Agulhas ring; and accumulation of sargassum inside of
a cold-core Gulf Stream ring.

Beyond the reach of the Maxey--Riley description is motion of
arbitrarily shaped objects; no known theory accounts for their
effects.  At the Maxey--Riley level there are terms and aspects
that we have ignored which may contribute to narrow the gap between
theory and observed motion.  One such term is the memory term, but
this only tends to slow down the inertial particle motion without
changing its qualitative dynamics fundamentally \citep{Daitche-Tel-11}.
Another neglected aspect is the dependence of fluid density on
spatial position and time.  Time varying density effects were
investigated previously in idealized settings and found to be of
importance \citep{Tanga-Provenzale-94}.  The observational possibility
to account for these effects is provided by satellite sensing of
sea surface temperature and salinity.  An additional aspect is the
effect of submesosocale perturbations on the mesoscale motions of
interest to us here.  These may be of fully ageostrophic and possibly
three-dimensional nature \citep{McWilliams-08a} or still be balanced
to leading order, and thus essentially two-dimensional and
incompressible \citep{Klein-Lapeyre-09}.  The latter is particularly
interesting as it opens the way to a potentially more accurate
observationally-based velocity representation when high-resolution
wide-swath altimetry becomes operational \citep{Fu-Ferrari-09}.
The only observational improvement over altimetry-derived velocities
may then be expected from the addition of an Ekman drift component
estimated from satellite scattometer wind measurements.  This
typically small correction is regularly included, but in such a way
as to match observed drifting buoy velocities \citep{Lagerloef-etal-99},
which is not consistent with our inertial particle approach.
 
We finally note that a larger sample of drifting buoys and floating
matter than that considered here is required to further validate
our predictions, possibly improved by the consideration of fluid
density variations and Ekman drift effects.

\begin{acknowledgments}
  Support for this work was provided by NOAA/AOML and UM/CIMAS, a
  grant from the BP/The Gulf of Mexico Research Initiative, and NSF
  grant CMG0825547.
\end{acknowledgments}

\appendix

\section{Reduced Maxey--Riley equation}\label{app:rmr}

The second-order ODE \eqref{eq:mr} is equivalent to the following
first-order ODE set:
\begin{equation}
  \dot{x} = v^\mathrm{p},\quad 
  \dot{v}^\mathrm{p} = f (\delta v - v^\mathrm{p})^\perp
  + \tau^{-1} (v - v^\mathrm{p}). 
  \label{eq:mr2}
\end{equation}
Taking $L$ and $L/V$ as length and time scales, respectively, and
making $\tau|f| = 1$ so that $\varepsilon =\tau V/L = \Ro \ll 1$,
the nondimensional form of \eqref{eq:mr2} reads:
\begin{equation}
  \dot{x} = v^\mathrm{p},\quad
   \varepsilon^2\dot{v}^\mathrm{p} =  \sign\!f\varepsilon
   (\delta v - v^\mathrm{p})^\perp + v - v^\mathrm{p}.
	\label{eq:nmr2}
\end{equation}
Inspection of \eqref{eq:nmr2} reveals that $x$ is a slow variable
that changes at $O(1)$ speed, while $v^\mathrm{p}$ is a fast variable
varying at $O(\varepsilon^{-2})$ speed.  Consequently, \eqref{eq:nmr2}
represents a singular perturbation problem.  To regularize it, we
displace and rescale time as $\varepsilon^{-2} (t - t_0)$.  Denoting
with a circle differentiation with respect to this fast time variable,
\eqref{eq:nmr2} transforms into
\begin{equation}
  \mathring{x} = \varepsilon^2 v^\mathrm{p},\quad
  \mathring{v}^\mathrm{p} =  \sign\!f\varepsilon (\delta
  v - v^\mathrm{p})^\perp + v - v^\mathrm{p},\quad
  \mathring{t} = \varepsilon^2.
  \label{eq:nmr3}
\end{equation}
The $\varepsilon = 0$ limit of system \eqref{eq:nmr3} has a
manifold of fixed points.  This manifold is normally attracting,
and hence survives for small $\varepsilon > 0$ in the form
\begin{equation}
  v^\mathrm{p} = v + \varepsilon v_1 + O(\varepsilon^2).
\end{equation}
Plugging this asymptotic series expansion into the right-hand-side
equation of system \eqref{eq:nmr3} and equating $O(\varepsilon)$
terms, it follows that
\begin{equation}
  v_1 = \sign\!f(\delta - 1) v^\perp.
\end{equation}
Inserting this expression in the left-hand-side equation of system
\eqref{eq:nmr3}, the inertial equation \eqref{eq:rmr} follows once
dimensional variables are recovered.

Particle dynamics governed by the inertial equation \eqref{eq:rmr}
evolve, over the finite-time interval of interest, on a two-dimensional
manifold, $\mathcal{M}_\varepsilon$, in the phase space with
coordinates $(x, v^\mathrm{p}, t)$.  This manifold is often referred
to as \emph{slow} because \eqref{eq:rmr} restricted to it is a
slowly varying system of the form $\mathring{x} = \varepsilon^2
v^\mathrm{p}\vert_{\mathcal{M}_\varepsilon} = \varepsilon^2 v +
\varepsilon^3 v_1 + O(\varepsilon^4)$. As shown in
\citet{Haller-Sapsis-08}, this slow manifold attracts all inertial
particle motions exponentially.

\section{Data}\label{app:data}

The altimetric SSH data employed in this paper consist of background
and perturbation components.  The background SSH component is steady,
given by a mean dynamic topography constructed from satellite
altimetry data, in-situ measurements, and a geoid model
\citep{Rio-Hernandez-04}.  The perturbation SSH component is
transient, given by altimetric SSH anomaly measurements provided
weekly on a 0.25$^{\circ}$-resolution longitude--latitude grid.
This perturbation component is referenced to a 20-year (1993--2012)
mean, obtained from the combined processing of data collected by
altimeters on the constellation of available satellites
\citep{LeTraon-etal-98}.  Mean dynamic topography and altimetry
data are distributed by AVISO at http://\allowbreak www.\allowbreak
aviso.\allowbreak oceanobs.\allowbreak com.

The RAFOS float trajectory data belong to the extensive dataset
constructed from float deployments in the California Undercurrent
over the period 1992--2010 \citep{Collins-etal-13}.  Acoustically
tracked, RAFOS floats are quasi-isobaric, with their density varying
with ambient temperature changes as a result of differing thermal
expansions of the glass hull and aluminum end cap of the floats
\citep{Rossby-etal-86}.  As opposed to seawater parcels, the floats
sink when they warm and rise when they cool \citep{Swift-Riser-94}.
The specific floats considered in this paper are shallow (300 db)
floats number 105 and 106, obtained from http://\allowbreak
www.\allowbreak oc.\allowbreak nps.\allowbreak edu/\allowbreak
npsRAFOS.

The surface drifters in the Loop Current ring were deployed from
air by the 53rd Hurricane Hunter Squadron ahead of hurricane Rita.
Equipped to monitor surface conditions, these drifters were of
Minimet (drogue at 15 m) and ADOS (with a 100-m-long thermistor
chain hanging below) types.  The trajectories of these drifters are
available from the NOAA Global Drifter Program at http://\allowbreak
www.\allowbreak aoml.\allowbreak noaa.\allowbreak gov/\allowbreak
phod/\allowbreak dac.

Three surface drifters were deployed in ring Astrid during the
MARE-1 cruise.  These were standard spherical WOCE/TOGA drifters,
fitted with an 8-m-long holey sock drogue at 15 m, with their
positions tracked using the Argos satellite system
\citep{Sybrandy-Niiler-91}.  The trajectories of these drifters are
not available from any database.  The trajectory of the drifter
considered in this paper was digitalized from Fig.\ 4 of
\citet{vanAken-etal-03} and spline fitted.  We have not been able
to reliably digitalize from this figure the other two trajectories,
which exhibit a qualitatively similar behavior.

Finally, the MERIS image shown in Fig.\ \ref{fig:sarga} is of L1b
Maximum Chlorophyll Index (MCI), a spectrometer parameter traditionally
used to detect and track \emph{Sargassum} \citep{Gower-King-08}.
MERIS imagery is available from ESA Earth Online at
https://earth.\allowbreak esa.\allowbreak int/\allowbreak web/\allowbreak
guest/\allowbreak data-access.

\section{Computational details}\label{app:numerics}

The flow maps associated with \eqref{eq:dxdt} and \eqref{eq:rmr}
were obtained from integration for initial positions on a regular
0.5-km-width grid covering the domain of interest.  This was done
using a stepsize-adapting fourth-order Runge--Kutta method with
interpolations obtained using a cubic scheme. The derivative of the
flow maps were computed using finite differences on an auxiliary
0.1-km-width grid of four points neighboring each point in the above
grid. Integrations of \eqref{eq:eta} and \eqref{eq:xi} were carried
out using the same method while enforcing a unique orientation for
the corresponding vector fields at each integration step (recall
that these are constructed from eigenvector fields, which are not
globally orientated). Detailed algorithmic steps for the extraction
of coherent Lagrangian eddies are outlined in \citet{Haller-Beron-13}.
The numerical computation of attracting iLCS involves the same
algorithmic steps as those for attracting LCS; these are outlined
in \citet{Haller-Beron-12} and \citet{Farazmand-Haller-13}.

%

\end{document}